\documentclass{article}
\usepackage{spconf,amsmath,graphicx}
\usepackage{amsfonts}
\usepackage{booktabs}
\usepackage{multirow}
\usepackage{url}
\usepackage{amssymb}
\usepackage{siunitx}
\usepackage{pifont}
\usepackage{cleveref}

\title{Text-to-Audio Grounding: Building Correspondence Between Captions and Sound Events}
\name{Xuenan Xu,
      Heinrich Dinkel,
      Mengyue Wu,
      Kai Yu\thanks{Mengyue Wu and Kai Yu are the corresponding authors.}}
\address{MoE Key Lab of Artificial Intelligence\\
SpeechLab, Department of Computer Science and Engineering\\
AI Institute, Shanghai Jiao Tong University, Shanghai, China\\
        \{\textit{wsntxxn, richman, mengyuewu, kai.yu}\}@sjtu.edu.cn\\ 
 }
\begin{document}
%
\maketitle

\begin{abstract}
Automated Audio Captioning is a cross-modal task, generating natural language descriptions to summarize the audio clips' sound events. 
However, grounding the actual sound events in the given audio based on its corresponding caption has not been investigated.
This paper contributes an \textit{AudioGrounding} dataset\footnote{\url{https://github.com/wsntxxn/TextToAudioGrounding}}, which provides the correspondence between sound events and the captions provided in Audiocaps, along with the location (timestamps) of each present sound event.
Based on such, we propose the text-to-audio grounding (TAG) task, which interactively considers the relationship between audio processing and language understanding. 
A baseline approach is provided, resulting in an event-F1 score of 28.3\% and a Polyphonic Sound Detection Score (PSDS) score of 14.7\%.
\end{abstract}
\section{Introduction}
\label{sec:intro}

Using natural language to summarise audio content, commonly referred as Automated Audio Captioning (AAC), has attracted much attention in recent studies~\cite{drossos2017automated,wu2019audio,koizumi2020_t1,wuyusong2020_t6}.
Compared with other audio processing tasks like Acoustic Scene Classification (ASC) and Sound Event Detection (SED), which aim to categorize audio into specific scenes or event labels, AAC allows the model to describe audio content in natural language, a much more unrestricted text form.
AAC can thus be seen as a less structural summarization of sound events.
However, the correspondence between sound event detection and natural language description is rarely investigated. 
To achieve human-like audio perception, a model should be able to generate an audio caption and understand natural language grounded in acoustic content, i.e., grounding (detecting) each sound event mentioned in a given audio caption to corresponding segments in that audio.
Explicit grounding of sound event phrases from the corresponding audio is key to audio-oriented language understanding. 
Moreover, it would be beneficial for generating captions with more accurate event illustrations and localized AAC evaluation methods.

Although such an audio grounding task (text-to-audio grounding, TAG) is relatively novel in audio understanding and audio-text cross-modal research, it is related to the following problems.

\textbf{Visual Grounding} A similar task to TAG is object grounding in Computer Vision (CV) using images or videos.
The \textit{Flickr30k Entities}~\cite{plummer2015flickr30k} is the first public dataset for image grounding. 
Image object grounding has become a research hotspot since then~\cite{rohrbach2016grounding,yeh2018unsupervised,akbari2019multi}.
Recently a plethora of work focus on new datasets and approaches for video object grounding~\cite{zhou2018weakly,chen2019object,zhou2019grounded}.
Like audio-text grounding, visual grounding requires a model to predict bounding boxes (2d coordinates) in an image or video frame for each object described in the caption.

\textbf{Sound Event Detection (SED)}
SED aims to classify and localize particular sound events in an audio clip.
With the growing influence of Detection and Classification of Acoustic Scenes and Events (DCASE) challenge~\cite{serizel2018large}, research interest in SED has soared recently.
TAG can be viewed as text-query-based SED, focusing on localizing sound events described by queries.
Due to SED and TAG's intrinsic correlation, we borrow common approaches and evaluation metrics from SED as a benchmark for TAG.

\begin{figure}[!htpb]
    \centering
    \includegraphics[width=\linewidth]{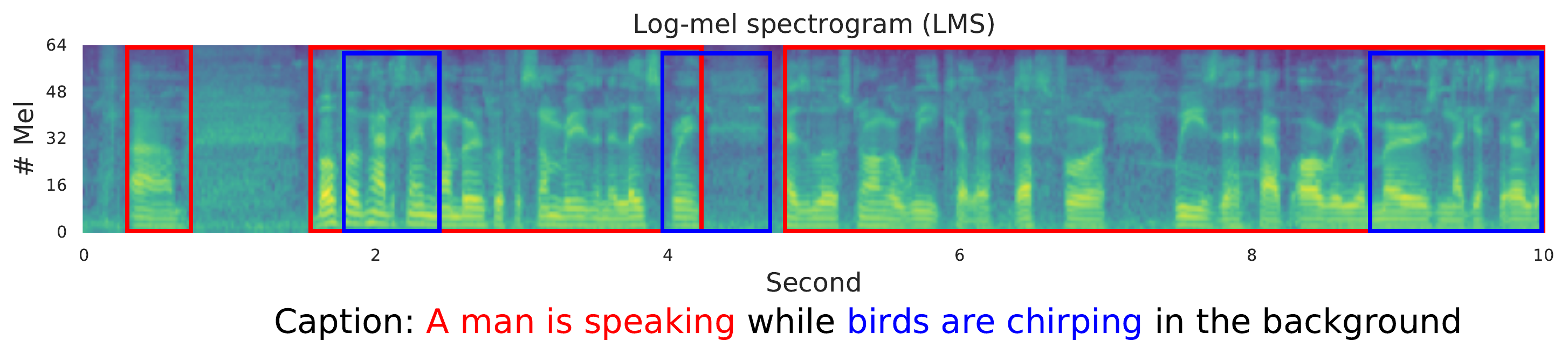}
    \caption{An example for \textit{TextToAudioGrounding}. For an audio clip and its corresponding caption, on- and off-set timestamps for each sound event phrase are provided. 
    In this example, both ``a man speaking'' (red) and ``birds chirping'' (blue) point to multiple segments (presented by rectangles in the figure).}
    \label{fig:data_sample}
\end{figure}

\begin{figure*}[!htpb]
    \centering
    \includegraphics[width=0.95\textwidth]{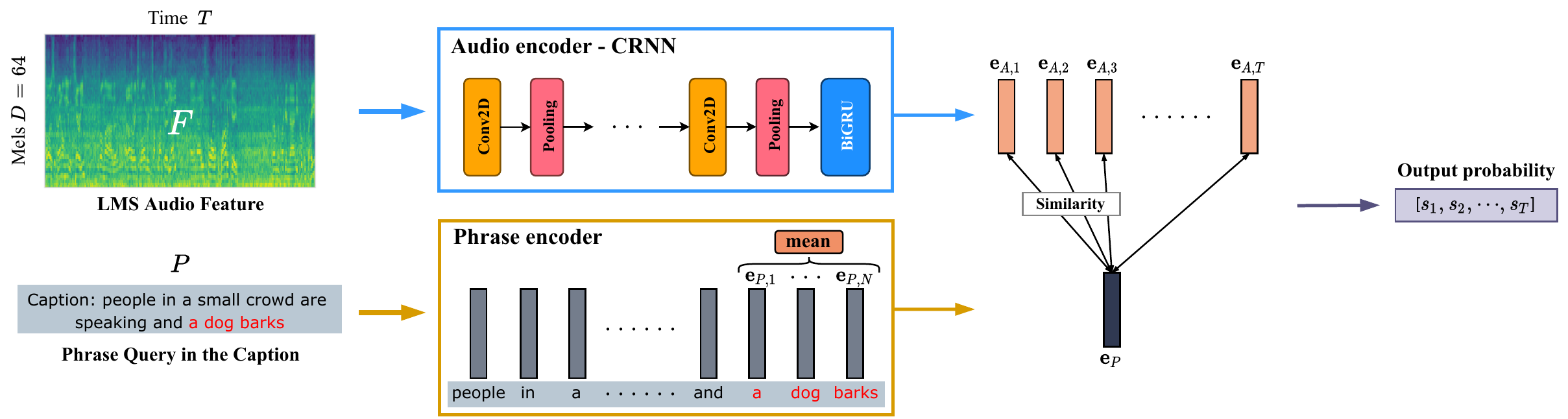}
    \caption{The proposed baseline model structure for TAG. A CRNN encoder outputs a sequence of audio embedding $\{\mathbf{e}_{A,t}\}_{t=1}^T$ from the LMS input $F \in \mathbb{R}^{T\times D}$. 
    The phrase query (containing $N$ words) is encoded into $\mathbf{e}_P$ by taking the mean of all word embeddings $\{\mathbf{e}_{P,n}\}_{n=1}^N$ in the query. 
    Prediction of the sound events' on- and off-sets are based on the similarity between $\{\mathbf{e}_{A,t}\}_{t=1}^T$, and $\mathbf{e}_{P}$.}
    \label{fig:model}
\end{figure*}

An audio grounding task inevitably consists of two parts. 
First is the extraction of sound event phrases from natural language caption, e.g., ``people speak'' and ``dogs bark'' can be obtained from the caption ``people speak while dogs bark''.
The second stage is concerned with traditional SED, detecting a sound event presence along with its onset and offset timestamps in the given audio clip. 
The prerequisite is a dataset that simultaneously provides audio, captions and the segmentation of sound events grounded from the caption.
To the best of our knowledge, no existing datasets or tasks are focusing on text-to-audio grounding.

We contribute \textit{AudioGrounding} dataset (\Cref{sec:dataset}) in this paper, providing a corresponding series of \textit{audio - caption - sound event phrase - sound event timestamp segmentation} to enable a more interactive cross-modal research within audio processing and natural language understanding. 
An illustration from \textit{AudioGrounding} is shown in \Cref{fig:data_sample}. 
With this dataset, we consider TAG, which localizes corresponding sound events in an audio clip from a given language description.
A baseline approach for the new TAG task is also proposed, see \Cref{sec:approach}.
\Cref{sec:experiments} details the experiment results and the analyses of such a TAG task, with conclusions provided in \Cref{sec:conclusion}.




\section{The Audio Grounding Dataset}
\label{sec:dataset}


Our \textit{AudioGrounding} dataset entails \num[group-separator={,}]{4994} audios, with one caption per audio in the training set, five captions per audio in the validation, and test sets. 
We provide caption-oriented sound event tagging for each audio, along with each sound event's segmentation timestamps. 
The audio sources are rooted in \textit{AudioSet}~\cite{gemmeke2017audio} and the captions are sourced from \textit{Audiocaps}~\cite{kim2019audiocaps}. 

\subsection{Audio and Caption Tailoring}
\textit{AudioSet} is a large-scale manually-annotated sound event dataset.
Each audio clip has a duration of up to ten seconds, containing at least one sound event label.
\textit{AudioSet} consists of a 527 event ontology, encompassing most everyday sounds.

\textit{Audiocaps}~\cite{kim2019audiocaps} is by far the largest AAC dataset, consisting of \num[group-separator={,}]{46000}+ audio clips ($\approx$ 127 hours) collected from \textit{AudioSet}. 
One human-annotated caption is provided for the training dataset while five captions for validation and test sets, respectively. 
Since the entire \textit{Audiocaps} dataset is a subset of \textit{AudioSet}, sound event labels can be obtained for each audio clip in \textit{Audiocaps}.

It should be noted that though \textit{AudioSet} provides sound tags and \textit{Audiocaps} consists of descriptive captions, there is no direct link between these two annotations. 
As we would like to enhance the diversity of the sound events included, we selectively choose audio clips with more than four sound tags, resulting in \num[group-separator={,}]{4994} audio clips sourced from \textit{Audiocaps}.
For a successful text-to-audio grounding, each audio clip should have not only a caption description (``A man is speaking while birds are chirping in the background''), but also the corresponding sound event phrases retrieved from the caption (``A man is speaking'', ``bird are chirping''), and the on- and off-sets of these sound events. 

\subsection{Annotation Process}
Our annotation process is decoupled into two stages: 
(1) sound event phrases are extracted automatically from captions; (2) we invite annotators to merge extracted phrases that correspond to the same sound event and provide the duration segmentation of each sound event.\\

\noindent\textit{A. Extracting Sound Event Phrases from Captions}

As mentioned above, the sound event labels provided in \textit{AudioSet} has no correspondence with the descriptive captions in \textit{Audiocaps}. 
Therefore we first extract sound event phrases from captions using NLTK~\cite{loper2002nltk}.
A phrase refers to a contiguous chunk of words in a caption.
Following standard chunking methods, we extract noun phrases (NP) and combinations of NP and verb phrases (NP + VP).
As sound descriptions usually stem from objects that sound (e.g., a cat) and verbs create the sound (e.g., meow), NP and NP + VP phrases can roughly summarize all possible sound events.\\

\noindent\textit{B. Phrase Merging and Segmentation}

Manual phrase merging is necessary since there might be repetitive and unwanted information in extracted phrases.  
For example, the caption in \Cref{fig:model} is chunked into three phrases: ``people'', ``a small crowd are speaking'' and ``a dog barks''.
However, ``people'' and ``a small crowd are speaking'' refer to the same sound event.
Based on the extracted phrases, annotators are required to label an audio clip in a two-step process:
\begin{enumerate}
    \item Merge phrases describing the same sound event into a single set and identify the number of sound events mentioned in the audio;
    \item Segment each sound event with the on- and off-set timestamps.
\end{enumerate}

\begin{figure}[!htpb]
    \centering
    \includegraphics[width=0.85\linewidth]{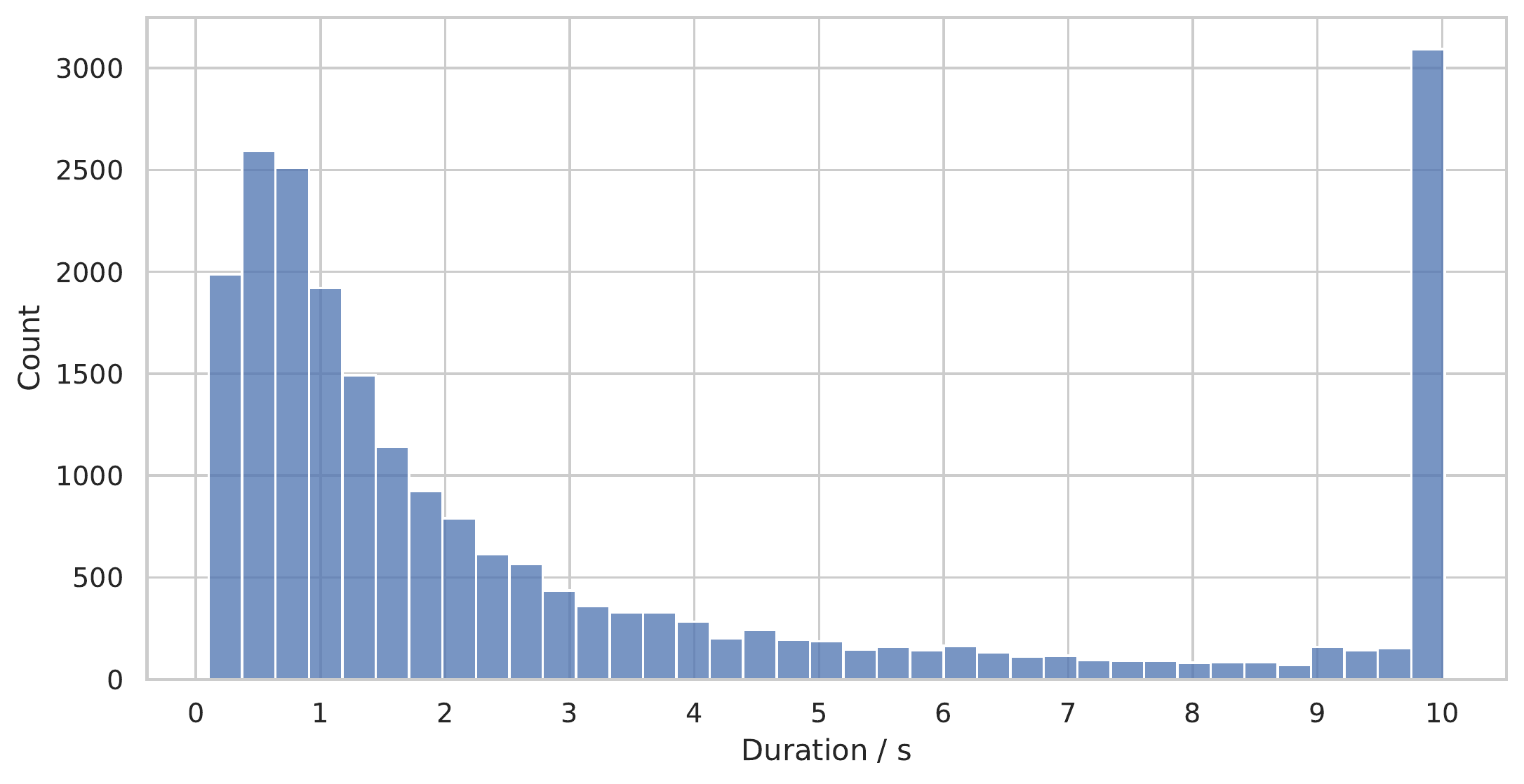}
    \caption{Duration distribution of annotated sound events mentioned in phrases within the proposed \textit{AudioGrounding} dataset.}
    \label{fig:duration_dist}
\end{figure}

\subsection{Data Description}
\label{subsec:data_description}

Our annotation results in a new audio-text grounding dataset: \textit{AudioGrounding}. 
It contains \num[group-separator={,}]{13985} corresponding sound event phrases and \num[group-separator={,}]{4994} captions (\textit{Audiocaps}).
After phrase merging, there are \num[group-separator={,}]{10910} sound events in total.
Sound events included are quite diversified, with the most frequent sound event (``a man speaks'') accounts for no more than 2\% of the dataset.
The sound event phrase duration distribution is shown in \Cref{fig:duration_dist}.
Most segments last for less than 2 s and the event phrases consist of several such short segments in a single audio clip, like speech, dog barking and cat meowing. 
However, a considerable proportion of events (e.g., wind, train) is present in the whole clip, lasting for almost 10 s.
We split the dataset according to the \textit{Audiocaps} setting, assigning each sample to the same subset (train/val/test) in \textit{Audiocaps}.
Detailed statistics are provided in \Cref{tab:datastatistics}.

\begin{table}[htpb]
    \centering
    \caption{Statistics of the \textit{AudioGrounding} Dataset.}
    \label{tab:datastatistics}
    \begin{tabular}{r||c|c|c}
    \toprule
     Split & \#Clips & \#Captions & \#Sound event phrases \\
     \midrule
    Train & 4489 & 4489 & 12373\\
    Val & 31 & 155 & 451\\
    Test & 70 & 350 & 1161\\
    \hline
    Total & 4590 & 4994 & 13958\\
    \bottomrule
    \end{tabular}
\end{table}

\vspace{-1mm}
\section{Text-to-Audio Grounding}
\label{sec:approach}

Since the primary motivation regards sound event grounding from phrases in audio captions, we use two separate encoders for audio and phrase query, respectively.
The input audio feature $F$ is encoded into an embedding sequence $\{\mathbf{e}_{A,t}\}_{t=1}^T$ while the query encoder outputs a phrase embedding $\mathbf{e}_P$ from the phrase query $P$ which consists of $N$ words.
Our baseline model architecture is illustrated in \Cref{fig:model}.
We apply $\text{exp}(-l2)$ as the similarity metric and binary cross-entropy (BCE) loss as the training criterion, following previous work in cross-modal audio/text retrieval~\cite{elizalde2019cross}. 
The similarity score between audio and phrase embedding $\mathbf{e}_{A, t}$ and $\mathbf{e}_{P}$ is calculated as:
{
\setlength{\abovedisplayskip}{3pt}
\setlength{\belowdisplayskip}{3pt}
\begin{equation}
    s_t = \text{sim} (\mathbf{e}_{A, t}, \mathbf{e}_{P}) = \exp(-\Vert \mathbf{e}_{A, t} - \mathbf{e}_{P} \Vert _2)
\end{equation}
}
During training, $\mathcal{L}_{\text{BCE}}$ between an audio-phrase pair is calculated as the mean of $\mathcal{L}_{\text{BCE}}$ between $\mathbf{e}_\text{A}$ at each frame $t$ and $\mathbf{e}_{P}$:
{
\setlength{\abovedisplayskip}{2pt}
\setlength{\belowdisplayskip}{2pt}
\begin{equation}
    \mathcal{L}_{\text{BCE}} = - \frac{1}{T}\sum_{t=1}^Ty_t\cdot\log(s_t) + (1 - y_t)\cdot\log(1 - s_t)
\end{equation}
}
where $y_t \in \{0,1\}$ is a strongly labeled indicator for each $t$. 
During evaluation, $\{s_t\}_{t=1}^T$ is transformed to $\{\hat{y}_t\}_{t=1}^T,\hat{y}_t \in \{0, 1\}$ by a threshold $\phi = 0.5$, representing the presence ($\hat{y}_t = 1, s_t > \phi$) or absence ($\hat{y}_t = 0, s_t \leq \phi$) of a phrase.\\

\vspace{-4mm}
\textbf{Audio Encoder} We adopt a convolutional recurrent neural network (CRNN)~\cite{dinkel2021towards} as the audio encoder.
The detailed CRNN architecture can be found in~\cite{xu2020crnn}.
It consists of five convolution blocks (with padded 3 $\times$ 3 convolutions) followed by a bidirectional gated recurrent unit (BiGRU).
L4-Norm subsampling layers are added between convolution blocks, reducing the temporal dimension by a factor of 4.
Finally, an upsampling operation is applied to ensure the output embedding has the same sequence length as the input feature.
The CRNN audio encoder outputs an embedding sequence $\{\mathbf{e}_{A,t}\}_{t=1}^T \in \mathbb{R}^{256}$.\\

\vspace{-4mm}
\textbf{Phrase Encoder}
For the phrase encoder, we only focus on extracting a representation for the phrase and leave out all other words in the caption.
The word embedding size is also set to 256 to match $\mathbf{e}_{A,t}$.
The mean of the word embeddings within a phrase is used as the representation:
{
\setlength{\abovedisplayskip}{3pt}
\setlength{\belowdisplayskip}{3pt}
\begin{equation}
    \mathbf{e}_{P} = \frac{1}{N}\sum_{n=1}^N \mathbf{e}_{P,n}
\end{equation}
}

\section{Experiments}
\label{sec:experiments}
\subsection{Experimental setup}
Standard Log Mel Spectrogram (LMS) is used as the audio feature since it is commonly utilized in SED.
We extract 64 dimensional LMS feature from a 40 ms window size and 20 ms window shift for each audio, resulting in $F \in \mathbb{R}^{T \times 64}$.
The model is trained for at most 100 epochs using the Adam optimization algorithm with an initial learning rate of 0.001.
The learning rate is reduced if the loss on the validation set does not improve for five epochs.
An early stop strategy with ten epochs is adopted in the training process.

\subsection{Evaluation}
Since TAG shares a similar target with SED, commonly used SED metrics are adopted for TAG evaluation. 
Specifically, we incorporate two metrics, being event-based metrics~\cite{mesaros2016metrics} and the newly proposed polyphonic sound detection score (PSDS)~\cite{bilen2020framework}. 
\begin{itemize}
    \item \textbf{Event-Based Metrics}  (Precision, Recall, F1) attach importance to the smoothness of the predicted segments, penalizing disjoint predictions. Regarding event-F1 scores, we set a t-collar value to 100 ms (due to large amounts of short events, see \Cref{fig:duration_dist}) as well as a tolerance of 20\% discrepancy between the reference and prediction duration for event-based metrics.
    \item \textbf{PSDS} is more robust to labelling subjectivity (e.g., to create one or two ground truths for two very close dog barks) and does not depend on operating points (e.g., thresholds). The default PSDS parameters are used~\cite{bilen2020framework}: $\rho_\text{DTC} = \rho_\text{GTC} = 0.5, \rho_\text{CTTC} = 0.3, \alpha_\text{CT} = \alpha_\text{ST} = 0.0, e_{max} = 100$.
\end{itemize}

Models achieving high scores in both event-based metrics and PSDS are expected to predict smooth segments while being robust to different operating points. 

\vspace{-3mm}

\begin{table}[!htpb]
    \centering
    \caption{Baseline TAG performance on the \textit{AudioGrounding} dataset. P, R, F1 represent the event-based precision, recall and, F1-score.} 
    \begin{tabular}{r||ccc|c}
    \toprule
    Model & $F_1$ & $P$ & $R$ & PSDS\\
    \midrule
    Random & 0.04 & 0.02 & 1.56 & 0.00\\
    Baseline & 28.30 & 28.60 & 27.90 & 14.70\\
    \bottomrule
    \end{tabular}
    \label{tab:result}
\end{table}

\vspace{-4mm}
\subsection{Results and Analyses}

\begin{figure}[htpb]
    \centering
    \includegraphics[width=\linewidth]{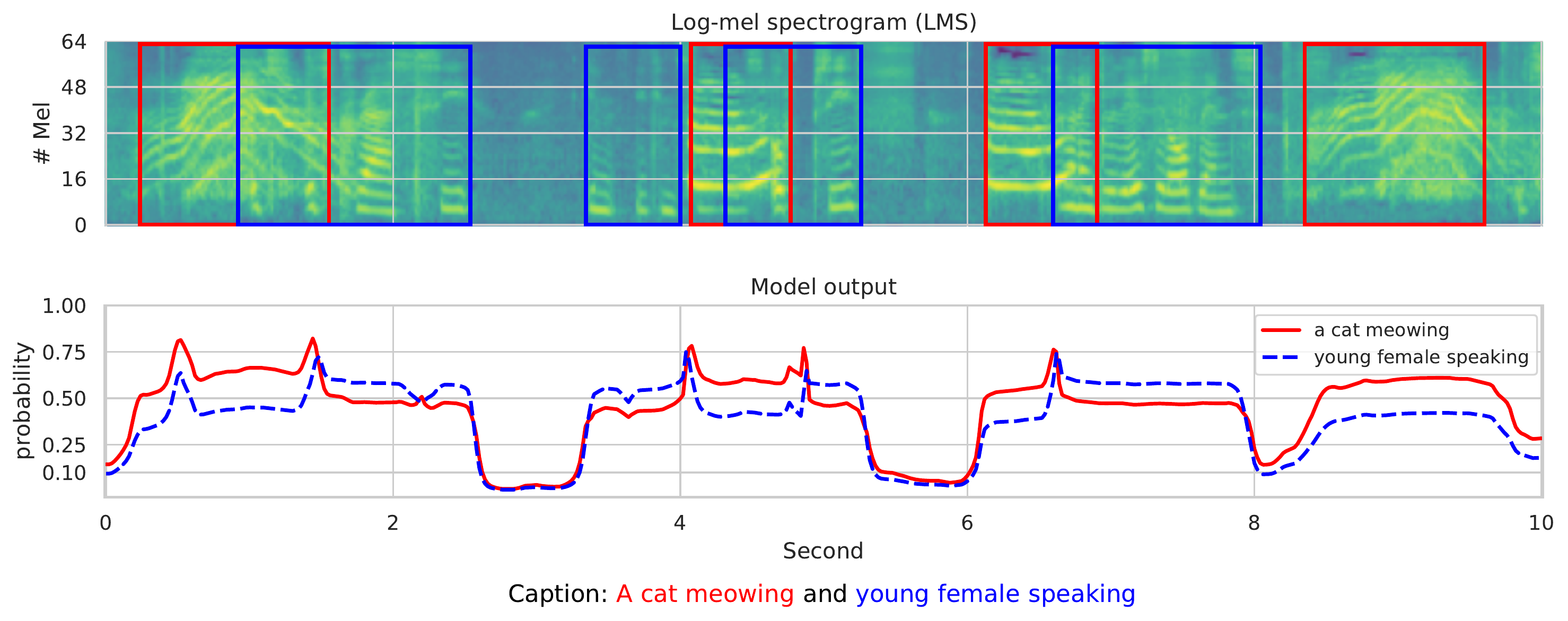}
    \caption{An example result of a TAG prediction on the \textit{AudioGrounding} dataset. The horizontal axis of the bottom figure denotes the output probability of a sound event according to the phrase query.}
    \label{fig:output_sample}
\end{figure}

We present the baseline TAG performance in \Cref{tab:result}.
The random guessing approach gives a random probability between 0 and 1 to each frame,  resulting in a 0.04\% event-F1 and 0.00\% PSDS, indicating the difficulty of this task.
In contrast, our proposed baseline model achieves 28.3\% event-F1 and 14.7\% PSDS, verifying its capability in audio and text understanding.
Despite the significant improvement against the random approach, we find that the baseline model tends to output high probability to salient parts of an audio clip, regardless of the phrase input.
An example is shown in \Cref{fig:output_sample}.
The output probabilities of both phrase inputs appear to be similar in their temporal distribution.
For the phrase query ``young female speaking'', the model assigns high presence probability to segments where either cats or female speech appear (e.g., the last two seconds).
This means the model only learns prominent audio patterns but neglects the information from the phrase query.
We change the phrase queries of each audio to a random phrase selected from all phrase queries of that audio.
After the modification, the event-F1 is still 19.6\%, indicating the insensitivity of our model to the phrase input.
Further research should be conducted on the text understanding and the fusion of these two modalities.

\vspace{-5mm}
\section{Conclusion}
\label{sec:conclusion}
In this paper, we propose a Text-to-Audio Grounding task to facilitate cross-modal learning between audio and natural language further. 
This paper contributes an \textit{AudioGrounding} dataset, which considers the correspondence between sound event phrases with the captions provided in Audiocaps~\cite{kim2019audiocaps} and provides the timestamps of each present sound event. 
A baseline approach that combines natural language and audio processing yields an event-F1 of 28.3\% and a PSDS of 14.7\%. 
We would like to explore better projection of audio and phrase embeddings as well as deeper interaction between these two modalities in future work. 
 
\section{Acknowledgement}
This work has been supported by National Natural Science Foundation of China (No.61901265), Shanghai Pujiang Program (No.19PJ1406300), and Startup Fund for Youngman Research at SJTU (No.19X100040009). Experiments have been carried out on the PI supercomputer at Shanghai Jiao Tong University.

\vfill\pagebreak

\bibliographystyle{IEEEtran}
\bibliography{refs}

\begin{thebibliography}{10}
\providecommand{\url}[1]{#1}
\csname url@samestyle\endcsname
\providecommand{\newblock}{\relax}
\providecommand{\bibinfo}[2]{#2}
\providecommand{\BIBentrySTDinterwordspacing}{\spaceskip=0pt\relax}
\providecommand{\BIBentryALTinterwordstretchfactor}{4}
\providecommand{\BIBentryALTinterwordspacing}{\spaceskip=\fontdimen2\font plus
\BIBentryALTinterwordstretchfactor\fontdimen3\font minus
  \fontdimen4\font\relax}
\providecommand{\BIBforeignlanguage}[2]{{%
\expandafter\ifx\csname l@#1\endcsname\relax
\typeout{** WARNING: IEEEtran.bst: No hyphenation pattern has been}%
\typeout{** loaded for the language `#1'. Using the pattern for}%
\typeout{** the default language instead.}%
\else
\language=\csname l@#1\endcsname
\fi
#2}}
\providecommand{\BIBdecl}{\relax}
\BIBdecl

\bibitem{drossos2017automated}
K.~Drossos, S.~Adavanne, and T.~Virtanen, ``Automated audio captioning with
  recurrent neural networks,'' in \emph{IEEE Workshop on Applications of Signal
  Processing to Audio and Acoustics (WASPAA)}.\hskip 1em plus 0.5em minus
  0.4em\relax IEEE, 2017, pp. 374--378.

\bibitem{wu2019audio}
M.~Wu, H.~Dinkel, and K.~Yu, ``Audio caption: Listen and tell,'' in \emph{Proc.
  IEEE International Conference on Acoustics, Speech and Signal Processing
  (ICASSP)}.\hskip 1em plus 0.5em minus 0.4em\relax IEEE, 2019, pp. 830--834.

\bibitem{koizumi2020_t1}
Y.~Koizumi, D.~Takeuchi, Y.~Ohishi, N.~Harada, and K.~Kashino, ``The {NTT}
  {DCASE2020} challenge task 6 system: Automated audio captioning with keywords
  and sentence length estimation,'' DCASE2020 Challenge, Tech. Rep., June 2020.

\bibitem{wuyusong2020_t6}
Y.~Wu, K.~Chen, Z.~Wang, X.~Zhang, F.~Nian, S.~Li, and X.~Shao, ``Audio
  captioning based on transformer and pre-training for 2020 {DCASE} audio
  captioning challenge,'' DCASE2020 Challenge, Tech. Rep., June 2020.

\bibitem{plummer2015flickr30k}
B.~A. Plummer, L.~Wang, C.~M. Cervantes, J.~C. Caicedo, J.~Hockenmaier, and
  S.~Lazebnik, ``Flickr30k entities: Collecting region-to-phrase
  correspondences for richer image-to-sentence models,'' in \emph{Proceedings
  of the IEEE International Conference on Computer Vision (ICCV)}, 2015, pp.
  2641--2649.

\bibitem{rohrbach2016grounding}
A.~Rohrbach, M.~Rohrbach, R.~Hu, T.~Darrell, and B.~Schiele, ``Grounding of
  textual phrases in images by reconstruction,'' in \emph{European Conference
  on Computer Vision (ECCV)}.\hskip 1em plus 0.5em minus 0.4em\relax Springer,
  2016, pp. 817--834.

\bibitem{yeh2018unsupervised}
R.~A. Yeh, M.~N. Do, and A.~G. Schwing, ``Unsupervised textual grounding:
  Linking words to image concepts,'' in \emph{Proceedings of the IEEE Computer
  Society Conference on Computer Vision and Pattern Recognition (CVPR)}, 2018,
  pp. 6125--6134.

\bibitem{akbari2019multi}
H.~Akbari, S.~Karaman, S.~Bhargava, B.~Chen, C.~Vondrick, and S.-F. Chang,
  ``Multi-level multimodal common semantic space for image-phrase grounding,''
  in \emph{Proceedings of the IEEE Computer Society Conference on Computer
  Vision and Pattern Recognition (CVPR)}, 2019, pp. 12\,476--12\,486.

\bibitem{zhou2018weakly}
L.~Zhou, N.~Louis, and J.~J. Corso, ``Weakly-supervised video object grounding
  from text by loss weighting and object interaction,'' in \emph{British
  Machine Vision Conference (BMVC)}, 2018, pp. 1--12.

\bibitem{chen2019object}
L.~Chen, M.~Zhai, J.~He, and G.~Mori, ``Object grounding via iterative context
  reasoning,'' in \emph{Proceedings of the IEEE International Conference on
  Computer Vision Workshops (ICCVW)}, 2019, pp. 1407--1415.

\bibitem{zhou2019grounded}
L.~Zhou, Y.~Kalantidis, X.~Chen, J.~J. Corso, and M.~Rohrbach, ``Grounded video
  description,'' in \emph{Proceedings of the IEEE Computer Society Conference
  on Computer Vision and Pattern Recognition (CVPR)}, 2019, pp. 6578--6587.

\bibitem{serizel2018large}
R.~Serizel, N.~Turpault, H.~Eghbal-Zadeh, and A.~P. Shah, ``Large-scale weakly
  labeled semi-supervised sound event detection in domestic environments,'' in
  \emph{Proceedings of the Detection and Classification of Acoustic Scenes and
  Events Workshop (DCASE)}, November 2018, pp. 19--23.

\bibitem{gemmeke2017audio}
J.~F. Gemmeke, D.~P. Ellis, D.~Freedman, A.~Jansen, W.~Lawrence, R.~C. Moore,
  M.~Plakal, and M.~Ritter, ``Audio set: An ontology and human-labeled dataset
  for audio events,'' in \emph{Proc. IEEE International Conference on
  Acoustics, Speech and Signal Processing (ICASSP)}.\hskip 1em plus 0.5em minus
  0.4em\relax IEEE, 2017, pp. 776--780.

\bibitem{kim2019audiocaps}
C.~D. Kim, B.~Kim, H.~Lee, and G.~Kim, ``Audiocaps: Generating captions for
  audios in the wild,'' in \emph{Proc. Conference of the North {A}merican
  Chapter of the Association for Computational Linguistics (NAACL)}, 2019, pp.
  119--132.

\bibitem{loper2002nltk}
E.~Loper and S.~Bird, ``Nltk: The natural language toolkit,'' in
  \emph{Proceedings of the ACL-02 Workshop on Effective Tools and Methodologies
  for Teaching Natural Language Processing and Computational Linguistics},
  2002, pp. 63--70.

\bibitem{elizalde2019cross}
B.~Elizalde, S.~Zarar, and B.~Raj, ``Cross modal audio search and retrieval
  with joint embeddings based on text and audio,'' in \emph{Proc. IEEE
  International Conference on Acoustics, Speech and Signal Processing
  (ICASSP)}.\hskip 1em plus 0.5em minus 0.4em\relax IEEE, 2019, pp. 4095--4099.

\bibitem{dinkel2021towards}
H.~Dinkel, M.~Wu, and K.~Yu, ``Towards duration robust weakly supervised sound
  event detection,'' \emph{IEEE Trans. Audio, Speech, Language Process.}, 2021.

\bibitem{xu2020crnn}
X.~Xu, H.~Dinkel, M.~Wu, and K.~Yu, ``A crnn-gru based reinforcement learning
  approach to audio captioning,'' in \emph{Proceedings of the Detection and
  Classification of Acoustic Scenes and Events Workshop (DCASE)}, Tokyo, Japan,
  November 2020, pp. 225--229.

\bibitem{mesaros2016metrics}
A.~Mesaros, T.~Heittola, and T.~Virtanen, ``Metrics for polyphonic sound event
  detection,'' \emph{Applied Sciences}, vol.~6, no.~6, p. 162, 2016.

\bibitem{bilen2020framework}
{\c{C}}.~Bilen, G.~Ferroni, F.~Tuveri, J.~Azcarreta, and S.~Krstulovi{\'c}, ``A
  framework for the robust evaluation of sound event detection,'' in
  \emph{Proc. IEEE International Conference on Acoustics, Speech and Signal
  Processing (ICASSP)}.\hskip 1em plus 0.5em minus 0.4em\relax IEEE, 2020, pp.
  61--65.

\end{thebibliography}

\end{document}